# Model Collapse: On Recursion, Noise, and Uncharted Machine Visions

Violaine Boutet de Monvel
violaine.boutetdemonvel@biblhertz.it[1]
Bibliotheca Hertziana – Max Planck Institute for Art History
MPRG Impett – *Machine Visual Culture: Artificial Intelligence and the History of Seeing*

**Abstract:**

Since 2023, computer scientists have warned against *model collapse* – the contamination of training sets with AI-generated outputs that progressively degrade model performance. Exemplifying a positive-feedback-driven failure, it produces effects such as word repetition or pixel noise, ultimately leading to a loss of meaning and coherence – at least from an engineering standpoint. From a creative one, however, collapse is not merely a breakdown: it also functions as a recursive mirror that recalls early analog video feedback experiments, raising once again the question of what happens when a system turns inward and sees itself. In such cases, so-called machine vision no longer transmits the world (as in *tele*-vision) but increasingly generates worlds from within. Drawing on media archaeology through case studies of both historical video synthesis techniques and contemporary artistic uses of machine learning, this paper examines what recursive training reveals about the dependent nature of AI-generated data. It argues that the potential effects of collapse challenge transhumanist ideals while inviting an aesthetic perspective, positioning noise and recursion as key concepts for understanding both artmaking and the AI ecosystem. Distributing agency across scales and networks, the latter currently remains reliant on new human-produced content, particularly within foundation models trained on massive datasets.

## Introduction: Art as Noise Magnification

The question of what model collapse reveals about the dependent nature of AI-generated data and its broader ecosystem must be situated within longer trajectories in the history of art, media, and cybernetics. My own point of entry arose from an attempt to define art in relation to engineering during my doctoral research at Sorbonne Nouvelle, titled *From Video Feedback to Generative AI: On Recursivity in the Arts and Media* (Boutet de Monvel, 2024). That project built an aesthetic bridge between early video art, from the 1960s to the 1980s, and contemporary practices associated with generative AI, which have gained momentum since the mid-2010s due to breakthroughs in deep learning, beginning with GANs (Goodfellow, et al., 2014) and soon after diffusion models (Sohl-Dickstein, et al., 2015). My aim was to reframe real-time analog synthesis – particularly video feedback techniques – as a precursor to current machine-learning-based art. To this end, I compared the room left for human mastery in the recursive operations of both closed-circuit video setups and generative AI models, which led me to articulate a networked, systemic view of artmaking that foregrounds technological agency alongside human intent, among other forces.

This approach culminated in a definition, which I first proposed in 2021 during a seminar at Sorbonne Nouvelle, and later published in *NECSUS*: *art strives to magnify the noise within and*

[1] Correspondence: violaine.bdm@gmail.com.

*via any given medium, as opposed to engineering that seeks to suppress it* (Boutet de Monvel, 2023, p. 121). In this context, *noise* refers to any agency that may introduce interference into the signal path beyond the artist's design – for example, the randomness of audience behavior in happenings, environmental contingencies like the weather, not to mention plain TV static, or the so-called hallucinations produced by generative AI. *Magnifying noise*, then, does not simply mean amplifying it – after all, engineers do that too when stress-testing or fine-tuning their systems through controlled injection (Igl, et al., 2019). It involves actively steering, sculpting, or orchestrating it, which does not necessarily imply ever fully controlling it. While not absolute, this distinction between engineering and art provided me with a gauge for analyzing and comparing a wider range of media practices, including structural film, vector synthesis, and glitch. More importantly, it proved crucial for understanding why recursive processes such as model collapse – like video feedback before it – could be approached not only as technical malfunctions but also as aesthetic opportunities.

The conceptual backdrop for my definition of art as noise magnification lies in information theory, specifically mathematician Claude E. Shannon's *Schematic diagram of a general communication system* (Shannon, 1948, p. 381). It describes a linear signal path from transmitter to receiver, potentially accumulating noise along the way – the latter, it should be noted, defined as an external input. Like the central cybernetic notion of feedback, in which information loops back to its source, Shannon's model was deeply influential on the seemingly polarized genealogy of avant-gardes from the early 1950s through the 1960s, including Fluxus, pop art, and minimalism. This period also saw the emergence of video art, thus providing the historical grounding for my PhD. Yet rather than pursuing the engineering goal of noise reduction, many artists deliberately worked against it: they explored ways to introduce contingency into their overall serial and/or systemic aesthetics to create difference within repetition. In the most radical cases, they even relinquished part of their agency to external factors, such as environmental or automated processes. Among these, *optical* and *system feedback* stand at the root of analog video synthesis, both constituting forms of recursive collapse in which so-called machine vision turns inward and no longer serves as a prosthetic extension of human perception.

Still contemplating noise magnification as a key dynamic in artmaking against the engineering imperative of minimizing uncertainty, this paper proposes to tackle model collapse from a media-archaeological perspective. The first part presents basic analog video synthesis techniques to highlight their formal resonance with later generative AI. The second examines how recursive training can degrade the performance of foundation models and at the same time produce effects that recall earlier artistic experiments with feedback. The third expands the frame to address the broader implications of AI's reliance on human-produced content, showing how model collapse challenges transhumanist ideals while opening onto actual, perhaps uncharted, machine visions.

## 1. Video Synthesis: Optical and Mixer Feedback

From a human-machine interaction standpoint, comparing past signal-based image production with today's generative AI is valuable because the former offers a situated, embodied perspective on synthetic processes that are otherwise difficult to grasp. These operations are now distributed across scales and networks, from training sets to latent spaces, and they unfold at speeds that far exceed human perception (Hansen, 2014, p. 4). Optical feedback, a technique at the forefront of analog video synthesis first explored for its aesthetic potential in the early 1960s, constitutes a striking counterpoint. Its basic setup simply involves pointing a camera at its own monitor – the visual equivalent of bringing a microphone close to its amplifier, whose resulting audio feedback is harsh to the ear and conventionally regarded as a malfunction. Yet the analogy ends here: in video, the outcome is far more hypnotic. When no tangible object intervenes, the continuous, abstracted flow of the apparatus looped back onto itself tends to evoke morphogenesis, such as the early stages of cellular differentiation, or even cosmogenesis. One of the earliest recorded applications is the 1963 opening title sequence of *Doctor Who: An Unearthly Child*, directed by Bernard Lodge for the BBC, where the effect was used to suggest intergalactic travel, specifically a wormhole (Figure 1).[2]

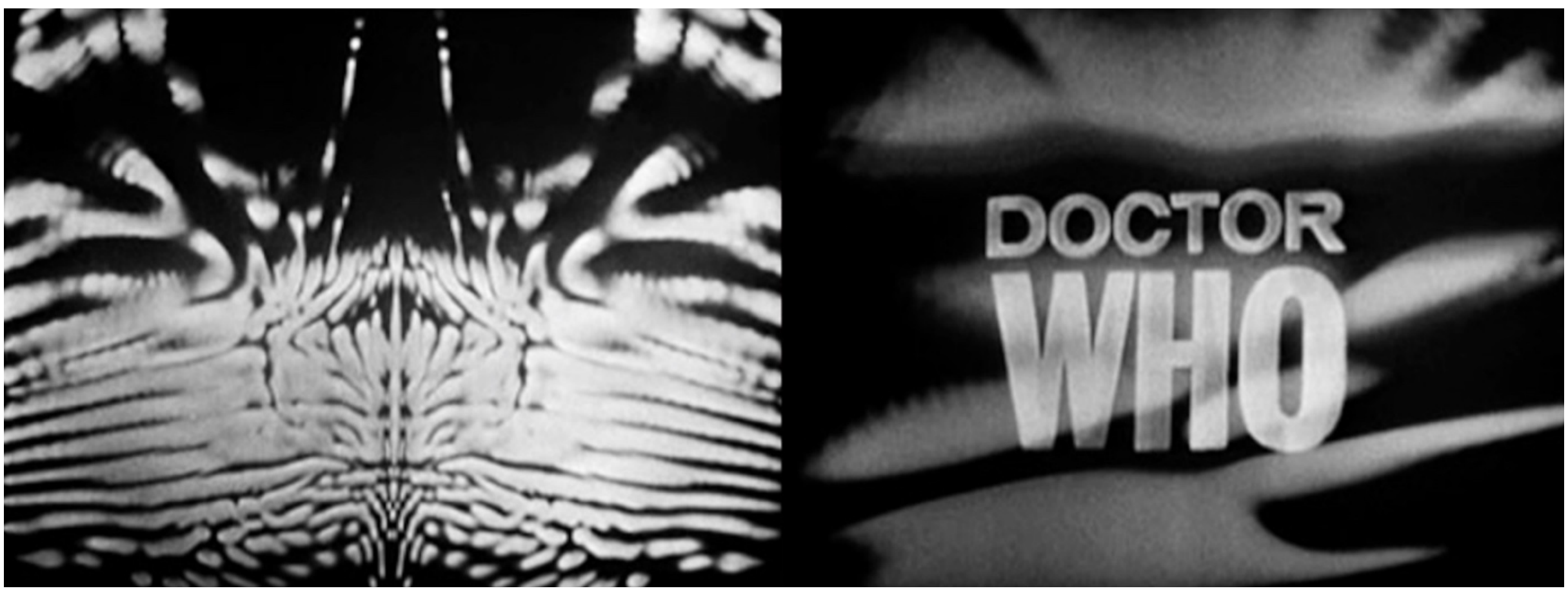

Figure 1: Bernard Lodge, *Doctor Who: An Unearthly Child*, opening title sequence, BBC, London, UK, 48 s, 1963. The images show optical feedback effects produced using standard broadcast equipment.

Such a basic video feedback setup – achievable, in practice, with any consumer device – can autonomously generate self-perpetuating inward abstractions, while further zooming in and pivoting the camera along the monitor's X and Y axes give rise to ever more intricate visual patterns. For instance, artist Peter Donebauer created his 1974 video *Entering* partly in this way (Figure 2).[3] Also produced for the BBC, it was made in real time using the standard equipment of the Royal College of Art Television Studio, in collaboration with electronic composer Simon Desorgher. The technique was then expertly employed to suggest "the experience of a foetus inside a mother's womb" (Donebauer, 1975, p. 30). Within the same morphogenetic realm, physicist James P. Crutchfield experimented extensively with this method as well during his doctoral research, defended in 1983 at the University of California, Santa Cruz, to study the emergence of order from chaos.[4] Video feedback constituted, in his view, "an almost ideal test bed from which to deepen and expand our understanding of spatial complexity and dynamic behavior" (Crutchfield, 1984, p. 229).

From the late 1960s onward, optical feedback also became a staple in the emerging music video genre, particularly through the hall-of-mirrors effect it produces when a tangible object is inserted into the loop. A notable example is the cascading "Magnifico" chorus in Queen's 1975 promotional video for *Bohemian Rhapsody*, directed by Bruce Gowers for the BBC's *Top of the Pops*.[5] In this case, the basic setup was expanded with a video mixer, allowing Freddy

---

[2] See Bernard Lodge, *Doctor Who: An Unearthly Child,* opening title sequence, BBC, London, UK, 48 s, 1963. Retrieved October 10, 2025, from https://youtu.be/sjDFvoRNpOM.

[3] See Peter Donebauer, *Entering,* video, BBC, London, UK, 6 min 50 s, 1974. Retrieved October 10, 2025, from https://archive.org/details/1974PeterDONEBAUERENTERING720.

[4] See James P. Crutchfield, *Space-Time Dynamics in Video Feedback,* video, Entropy Productions, Santa Cruz, USA, 15 min 33 s, 1984. Retrieved October 10, 2025, from https://youtu.be/B4Kn3djJMCE.

[5] See Bruce Gowers, *Bohemian Rhapsody*, promotional video, Elstree Studios, Borehamwood, UK, 6 min, 1975. Retrieved October 10, 2025, from https://youtu.be/fJ9rUzIMcZQ.

Mercury's figure to be chroma-keyed into the collapsing imagery without revealing the apparatus.

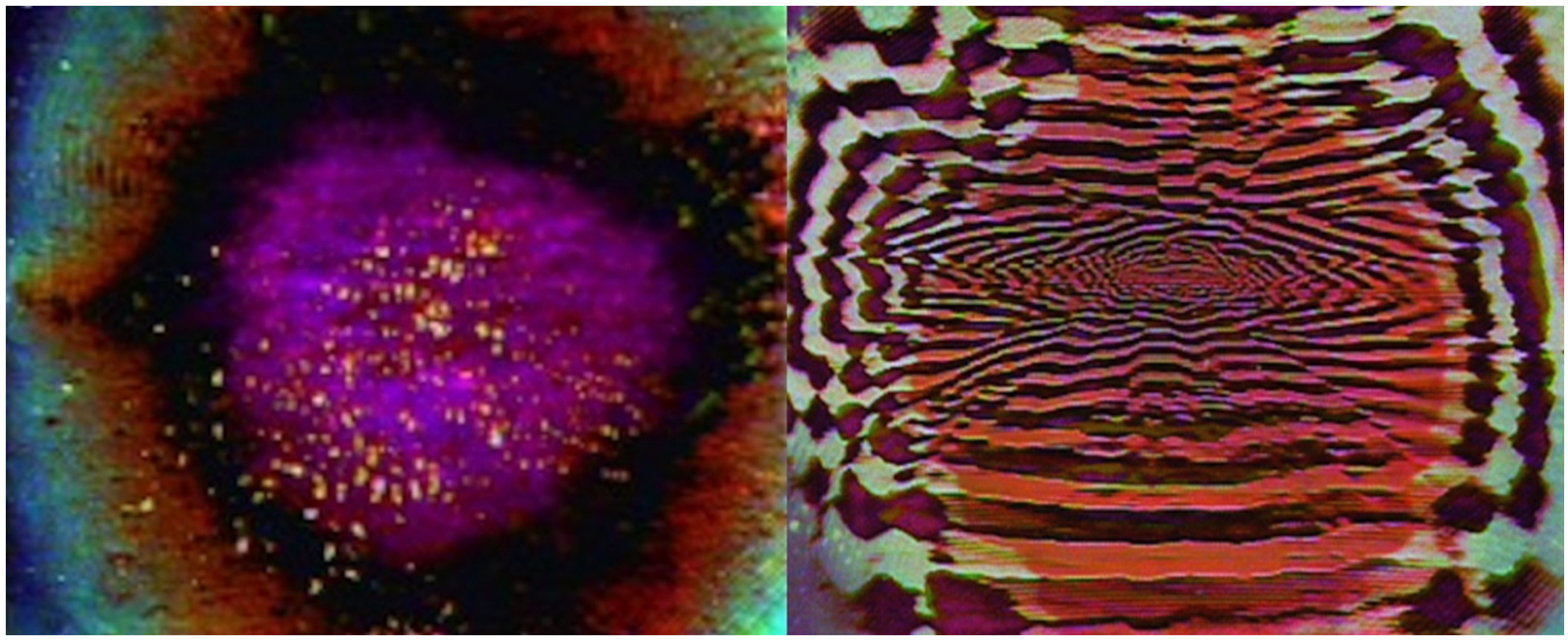

Figure 2: Peter Donebauer, *Entering*, video, BBC, London, UK, 6 min 50 s, 1974. The images show optical feedback effects produced using standard broadcast equipment.

This introduces the other major technique in video synthesis: system feedback, which – unlike optical feedback – requires no camera, only a monitor connected to a real-time image processing tool, such as a video synthesizer or a video mixer. The former, state-of-the-art device is equipped with an oscillator and sometimes a waveform generator, enabling it to produce a signal independently of any external referent. To achieve this with a standard video mixer, normally used to input an image (live or pre-recorded) for chroma-keying, the output must instead be immediately fed back in, thus creating a self-referential loop. Without further human intervention, this setup typically produces an RGB flow moving from left to right across the screen, while different models – from analog to digital – yield distinct visual artifacts. To give a recent example, and to show that this creative territory remains fertile to this day, Paloma Kop and Andrei Jay from the now-Phase Shift collective (previously Phase Space) realized their entire 2019 *WOODWAVEWORMROSE* using a hybrid WJ-MX50 video mixer, originally developed by Panasonic in 1991 (Figure 3).[6]

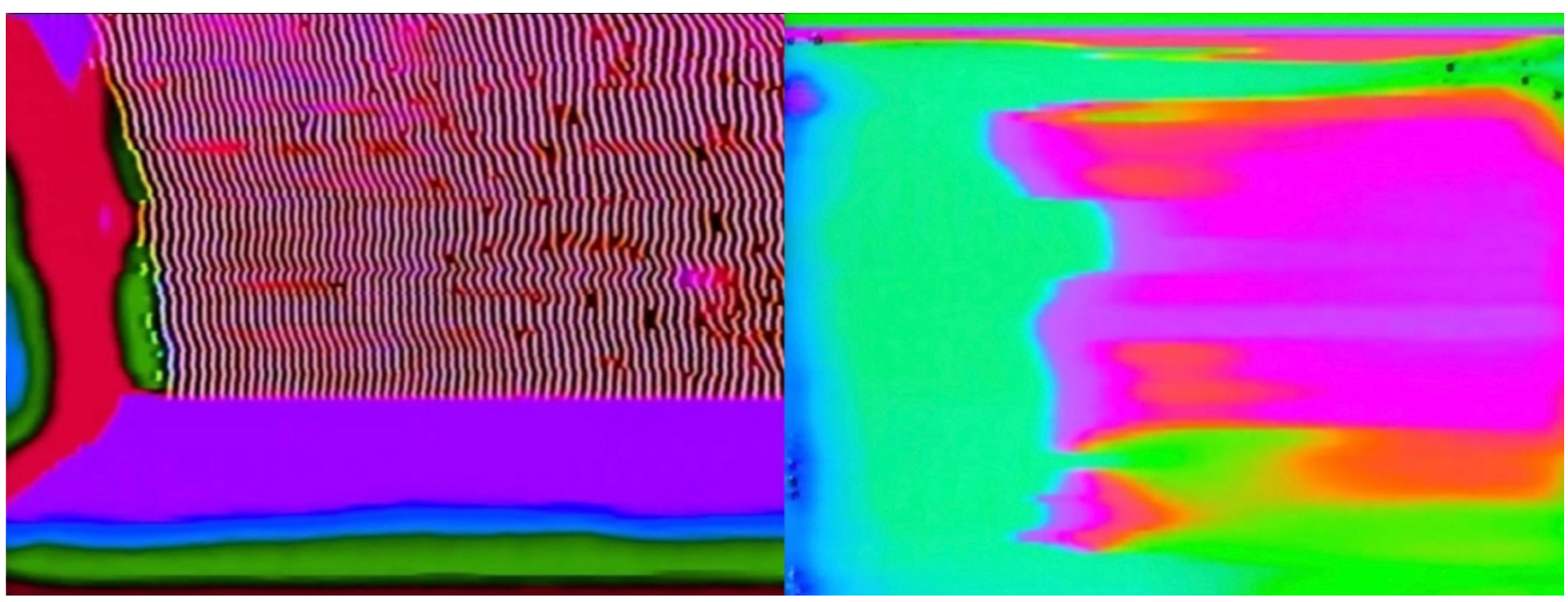

Figure 3: Paloma Kop and Andrei Jay, *WOODWAVEWORMROSE*, video, 15 min 47 s, 2019. The images show system feedback effects produced with a hybrid Panasonic WJ-MX50 video mixer model. Courtesy of the artists.

[6] See Paloma Kop and Andrei Jay, *WOODWAVEWORMROSE,* video, 15 min 47 s, 2019. Retrieved October 10, 2025, from https://youtu.be/WgbscQ34rBk.

Both optical and system feedback techniques – which can, of course, be combined (Gwin, 1972, p. 5) – have the power to generate, in real-time, an infinite, complex, and ever-changing *mise en abyme* of the external setup and/or its internal circuitry. They produce self-sustaining machine visions that, paradoxically, appear life-like precisely because they maintain acute sensitivity to interference, including further human intervention. An artist's touch, for example, manifests through the manipulation of knobs and sliders, prospectively adjusting parameters to magnify the signal's appearance on screen: its shape, color, brightness, contrast, motion, or rhythm. As with chaos theory and its butterfly effect (Lorenz, 1973), the slightest change in the initial setup triggers substantial, highly unpredictable shifts. Without factoring in the possible use of more advanced video synthesizers, such instability renders mixer feedback considerably trickier to classify than optical feedback, whose hall-of-mirrors effects – even when abstracted – lend themselves more readily to image recognition and comparative analysis across media. Setting this aside, another key distinction remains: system feedback requires no external reference, camera view, or pre-recorded material, which is central to understanding the aesthetic potential of model collapse, as well as the implications of this phenomenon for the broader AI ecosystem.

## 2. Model Collapse: Positive Feedback, Spiraling Entropy

In my PhD, I adopted a media-archaeological approach to show how earlier techniques of video or even vector synthesis could shed light on the recursive operations of generative AI. My definition of art as noise magnification naturally led to wonder whether an equivalent of mixer feedback might exist in diffusion-based text-to-image models. The question was less pressing in the case of GANs, whose very architecture is built on an explicit adversarial feedback loop between two artificial neural networks: one the one hand, a generator that synthesizes images from random pixel noise; on the other, a discriminator that distinguishes these samples from the original training data (Somaini, 2023, p. 92). An early example of noise magnification within this framework is Trevor Paglen's 2017 series *Adversarially Evolved Hallucinations*, in which the artist interrupted the machine-learning process before completion to produce uncanny images that hover at the threshold of abstraction, yet whose objects are already faintly recognizable (Paglen, 2020).

Early diffusion-based text-to-image models, beginning with OpenAI's DALL·E in 2021, did not allow users to explore the underlying architecture or construct custom training sets. It was only later in 2022, with Stability AI's open-source release of Stable Diffusion, that it became possible to intervene into the process, thereby magnifying potential noise. This raised a crucial question: what would happen if the logic of the *camera obscura* were broken – the pinhole principle that loosely underpins machine learning's tether to perceived reality, or more precisely, to our shared, digitized, or digital-native imagination? At its core, this meant asking: what might emerge if a model is trained not on external data, but recursively on its own outputs?

The answer soon pointed to collapse, a potential crisis in generative AI first discussed in 2023 and, to date, observed chiefly in experimental settings: training sets contaminated with synthetic data progressively degrade model performance, introducing noise. This phenomenon highlights a fundamental property of deep learning: the tendency to form positive feedback loops, which spiral toward entropy when AI-generated outputs begin to outweigh human content. To clarify, cybernetics distinguishes between two types of recursion (Wiener, 1948). Negative feedback – or homeostasis – reduces contingent noise and stabilizes conditions in a communication channel, whether living or non-living. In this sense, making a phone call audible was the primary goal of information theory, just as delivering a coherent, plausible output is that of deep learning.

By contrast, positive feedback amplifies noise, often with destructive consequences – death, for example, when the body can no longer regulate itself. Yet, as noted earlier, positive feedback may also be harnessed creatively to produce difference within repetition. I have even argued, in my PhD, that it constitutes the *condition sine qua non* for the renewal and diversification of art forms against the engineering imperative to minimize uncertainty, formalized as: *regenerated art / message (output) = original art / signal (input) + noise* (Boutet de Monvel, 2024, p. 60). After all, higher entropy corresponds to a greater amount of information, which in turn enables surprise and variation.

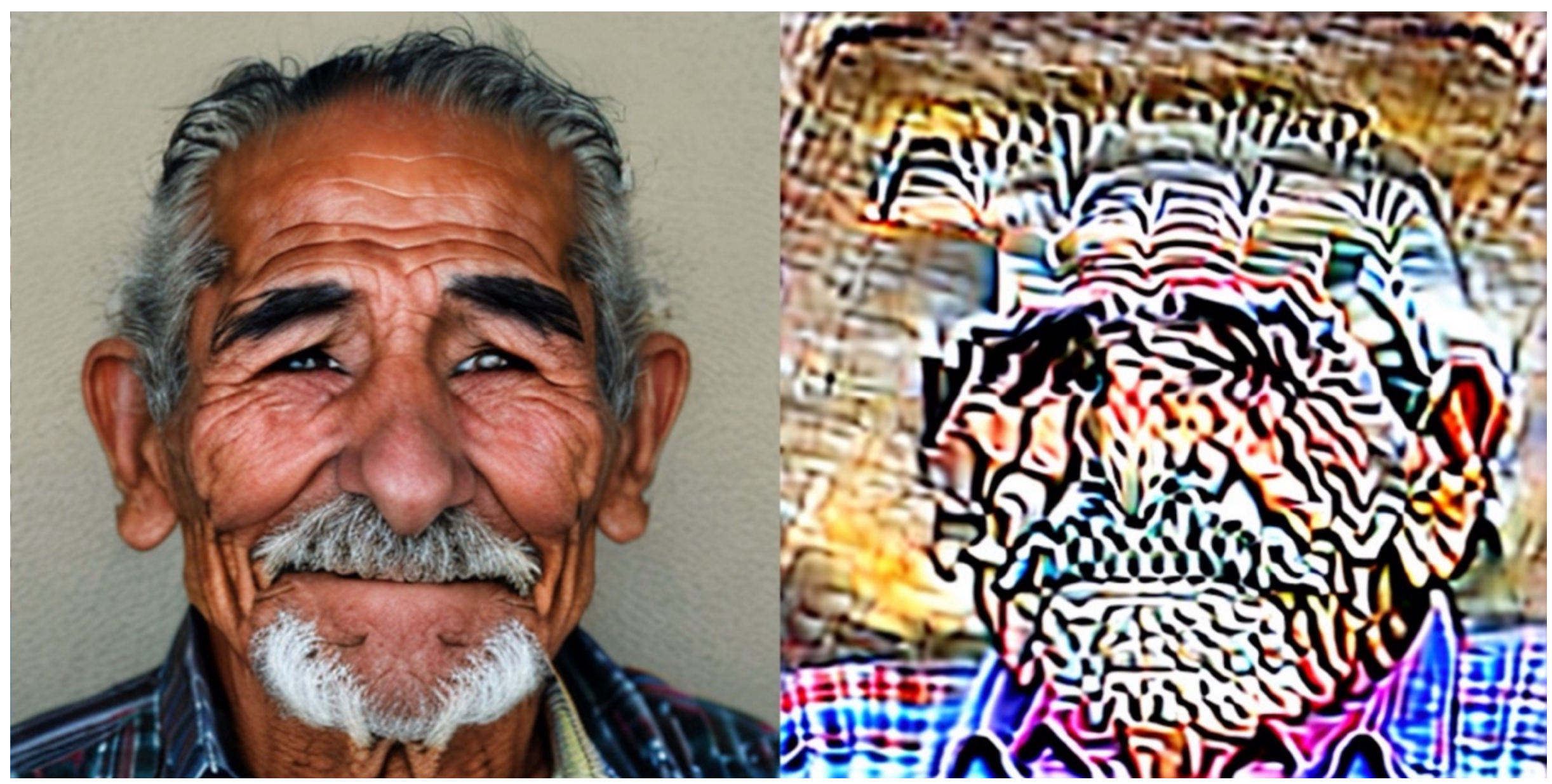

Figure 4: Matyás Boháček and Hany Farid, research data, 2023. The images show Stable Diffusion outputs after the first and sixth iterations of recursive training, respectively, starting from a 50% corrupted dataset and the prompt "older hispanic man."

One of the most comprehensive studies on model collapse, led by Ilia Shumailov at Oxford, defined it as a degenerative process whereby synthetic data "end up polluting the training set of the next generation [which] then mis-perceive[s] reality" (Shumailov, et al., 2024, p. 755). Assuming that AI-generated outputs already surpass the volume of human-produced content online, this issue arises directly from the introduction of LLMs in the late 2010s and poses potential risks for large-scale training if not carefully monitored. Specifically, Shumailov tested the short- and long-term effects of such collapse in GPT-3, launched by OpenAI in 2020. Over several weeks, the model was recursively trained on datasets increasingly contaminated by outputs from its prior synthetic cycles, resulting in word repetition, token breakdown, and a gradual loss of meaning.

Similar experiments were also conducted on text-to-image models earlier in 2023. At Stanford, Matyás Boháček and Hany Farid used Stable Diffusion to generate successive training sets from a single prompt: "older hispanic man" (Boháček, Farid, 2023, p. 2). With each iteration, the resulting images visibly degraded, accumulating pixel noise (Figure 4). Another team at Stanford, led by Sina Alemohammad, likened model collapse to *autophagy* – a biological term describing a homeostatic process in which a diseased cell recycles its own components (Alemohammad, et al., 2023, p. 1) – while ethology offers an even starker metaphor: *data cannibalism* (Pedersen, 2023). Viewed through the lens of my definition of art as noise magnification, however, phenomena deemed problematic from an engineering standpoint may instead present creative potential.

Ègor Kraft's 2023 recursive video *One and Infinite Chairs* provides an early example of artistic engagement with model collapse. It was inspired by Joseph Kosuth's 1965 conceptual assemblage *One and Three Chairs* – in which a wooden chair, its photograph, and a textual definition of the object were juxtaposed on a gallery wall, effectively prefiguring the generative AI framework. Kraft's remediation was created with Stable Diffusion, refeeding each generation of outputs into the next training set, starting from the prompt "a single chair on a white background" (Figure 5). [7] After six cycles, all trace of a chair had vanished, replaced by abstract, noisy forms evocative of mixer feedback, with no further human intervention. What began as a referential image eventually became a self-contained synthetic flow, suspended and adrift.

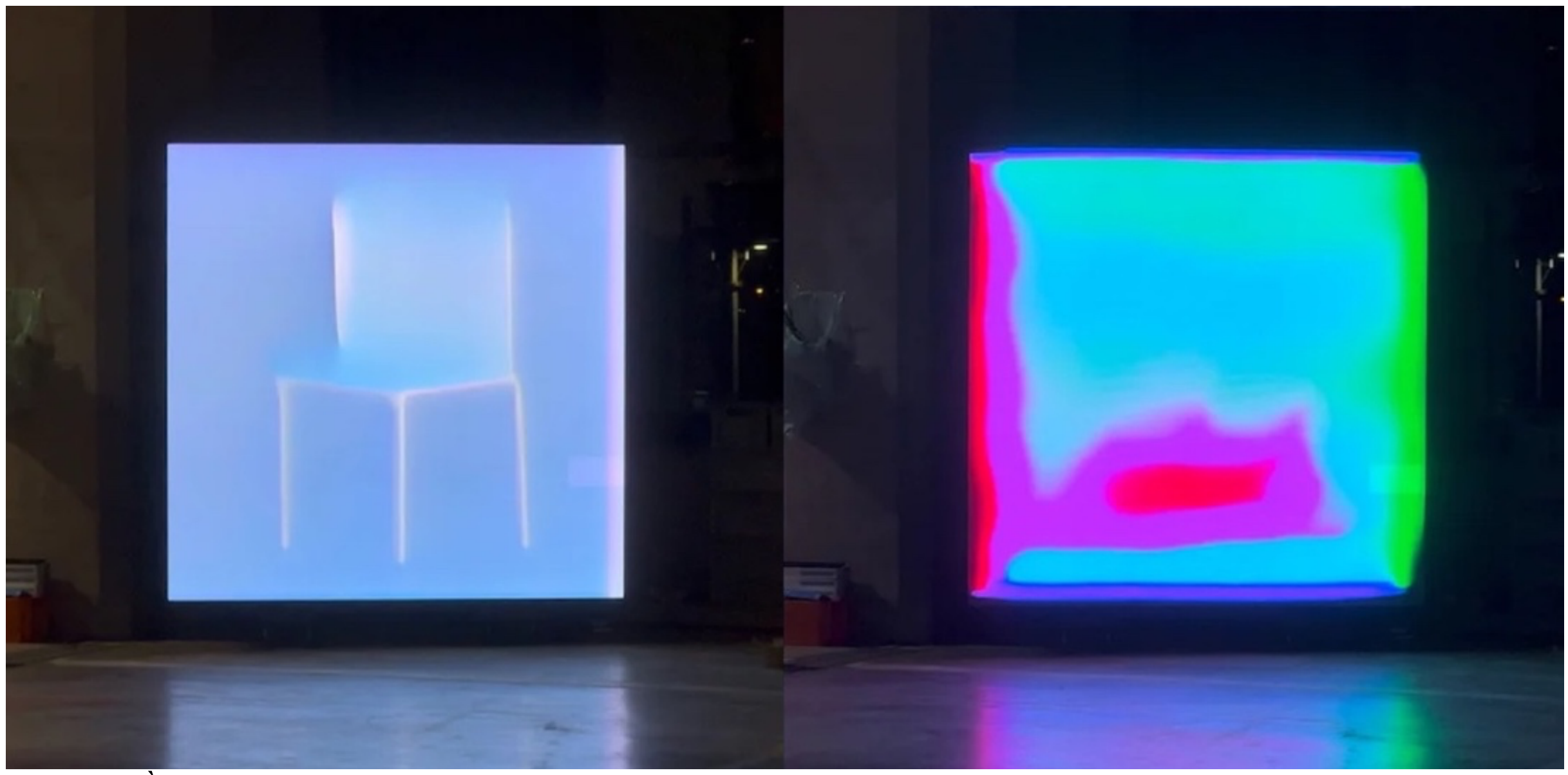

Figure 5: Ègor Kraft, *One and Infinite Chairs*, installation view, Postcity, Linz, Austria, 2023. The images show Stable Diffusion outputs after the third and sixth iteration of recursive training, respectively, starting from the prompt "a single chair on a white background." Courtesy of the artist.

## 3. Data Dependency: A Transhumanist Dystopia

In both LLMs and diffusion-based text-to-image models, recursive training thus produces word repetition or pixel noise, ultimately leading to a loss of meaning and coherence – at least from an engineering standpoint. Yet recognizing this bias is crucial for understanding model collapse – like video feedback before it – not merely as a technical failure but as an aesthetic and epistemological opportunity: a lens through which to rethink artmaking in general, and creative practice with generative AI in particular. Model collapse is, first and foremost, a recursive mirror that raises once again the question of what happens when a system turns inward and sees itself. In such cases, so-called machine vision no longer transmits the world (as in *tele*-vision) but increasingly generates worlds from within.

Before pursuing this speculative perspective, the underlying engineering issue behind this potential crisis must be emphasized. Model collapse is not simply about suppressing undesirable internal noise arising from synthetic contamination of training sets; it reflects a deeper data dependency. Generative AI seemingly produces infinite outputs, which should ideally shift discussions about creativity toward broader scalar dynamics beyond sole human agency (Krysztoforska, Kenny, 2024). In practice, however, the phenomenon demonstrates the opposite: current frontier models cannot sustain meaningful variation indefinitely without regular access to substantial amounts of fresh human material. A case in point is Meta, which in spring 2025 asked European Facebook users to consent to their posts being scraped to train MetaAI (Davies, 2025). This does not merely constitute another instance of surveillance capitalism; it points out to a possibly more disquieting reality: there may no longer be sufficient legally available human data online to support large-scale AI training. Although foundation models are pretrained offline, they still require periodic fine-tuning to remain effective, and this process depends on "the richness and variety that comes with human-generated content" (Gibney, 2024, p. 18) – in other words, external noise. What, then, does this data dependency reveal about the broader AI ecosystem?

[7] See Ègor Kraft, *One and Infinite Chairs*, film-essay trailer, 6 min 36 s, 2023. Retrieved October 10, 2025, from https://kraft.studio/chair/.

A transhumanist dystopia imagined by Grégory Chatonsky as early as 2022 offers a partial answer. Titled *Disnovation*, the work took the form of a multimedia installation speculating on humanity's fate had it survived its own extinction – a threat to which the current extractive and exploitative infrastructures of generative AI undoubtedly contribute. The ensemble featured several videos of recognizable faces, beginning with the artist himself at three life stages: adolescence, adulthood, and old age. These avatars were created using Unreal Engine, a software developed in 1998 by Epic Games to produce photorealistic interactive graphics without coding. The speech, meanwhile, was generated with a specially modified version of GPT-2, originally released by OpenAI in 2019. Chatonsky retrained the model on carefully selected textual datasets to deliver a nonsensical monologue, parodying the rhetorical tropes of innovation, self-help, and meditation.[8] Through this seemingly endless gibberish, the so-called disnovation was only vaguely described – an invention of unclear nature, ultimately destined to devastate the Earth.

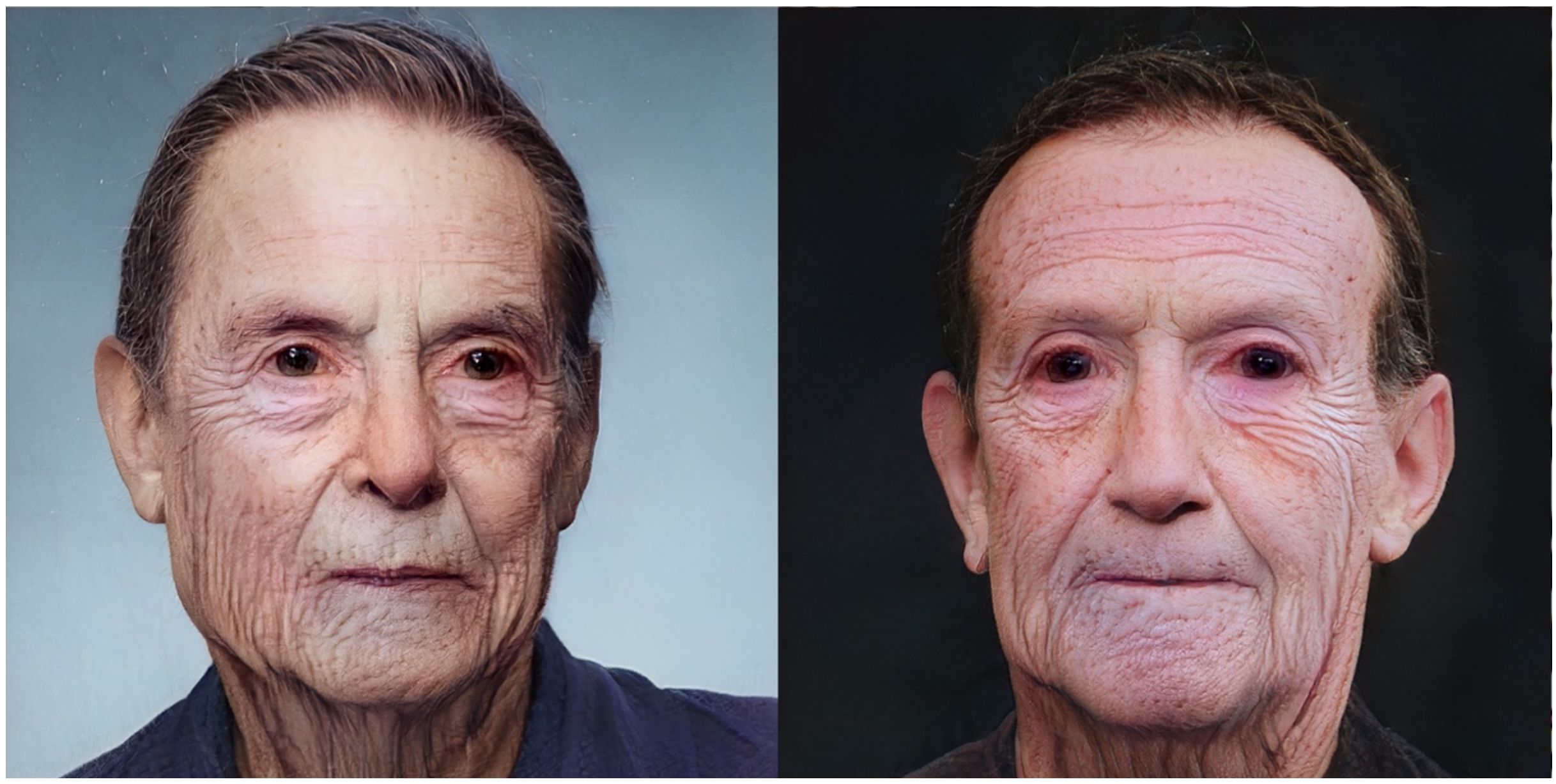

Figure 6: Grégory Chatonsky, *Disnovation*, installation details, 2022. The images show video portraits of Elon Musk and Mark Zuckerberg produced with Unreal Engine. Courtesy of the artist.

[8] See Grégory Chatonsky, *Disnovation v.1*, video documentation, 32 min 46 s, 2022. Retrieved October 10, 2025, from https://youtu.be/tYZOXVimcN0.

Beyond the artist's prophetic figure, the installation also included ten silent portraits representing GAFAM leaders – including Mark Zuckerberg, Jeff Bezos, and Elon Musk, the respective heads of Meta, Amazon, and Tesla (Figure 6). Their appearances had been drastically aged. According to the project synopsis, these tech magnates had fled to Mars centuries earlier and uploaded their memories into AIs, enabling their avatars to survive biological death (Chatonsky, 2022). Over time, these algorithmic reincarnations continued to communicate but no longer understood the original thoughts used to train them. Instead, they mechanically repeated phrases once spoken, like a form of prayer. After countless iterations, their speech lost all meaning, eventually fading into silence.

Aside from its dystopian tone, this fable engages with contemporary transhumanist ideology, which champions innovation at all costs – not merely to improve human life but to transcend its natural limits. While the ecological crisis remains one of humanity's most pressing challenges, this utopia has become increasingly entangled, since the turn of the 21st century, with a renewed space race – one that leading actors in the AI industry are actively supporting for profit (Crawford, 2021, p. 231). Among them, Bezos founded Blue Origin in 2000 and Musk SpaceX in 2002. Both aim to develop space tourism for a billionaire clientele and plan inhabited missions to the Moon by the end of the 2020s, as part of NASA's Commercial Lunar Payload Services (CLPS) and Artemis programs – potentially offering, if not a solution, at least an escape from Earth's ongoing devastation, albeit for a highly privileged few.

## Conclusion: AI and the Rhetoric of Space Exploration

To return to the muteness that ironically afflicted the avatars of transhumans in Chatonsky's recursive dystopia: its cause was, in fact, model collapse – before the term had been coined and the phenomenon formally studied by computer scientists. As discussed earlier, however, were it to occur significantly in real-world settings rather than under controlled laboratory conditions, it would be unlikely to result in silence. Instead, it might synthesize a kind of *glossolalia* – a speaking in tongues in which words disintegrate into tokens looping endlessly, unmoored from meaning. It could even give rise to a future language: a poetic, statistical equivalent of what *musique concrète* once pioneered, filled with novel syntactic artifacts or incongruities such as "arafed." This is precisely the terrain explored by Ting-Chun Liu and Leon-Etienne Kühr, who have created feedback loops within BLIP-2, an image-captioning model that generates descriptions from visual data (Liu, Kühr, 2025). The duo has also tested methods for collapsing Stable Diffusion without extended retraining or even textual prompting. Their experiments show that simply refeeding the model's outputs as inputs – beginning from a single image – rapidly leads to performance degradation and pixel noise (Liu, Kühr, 2024, Figure 7).

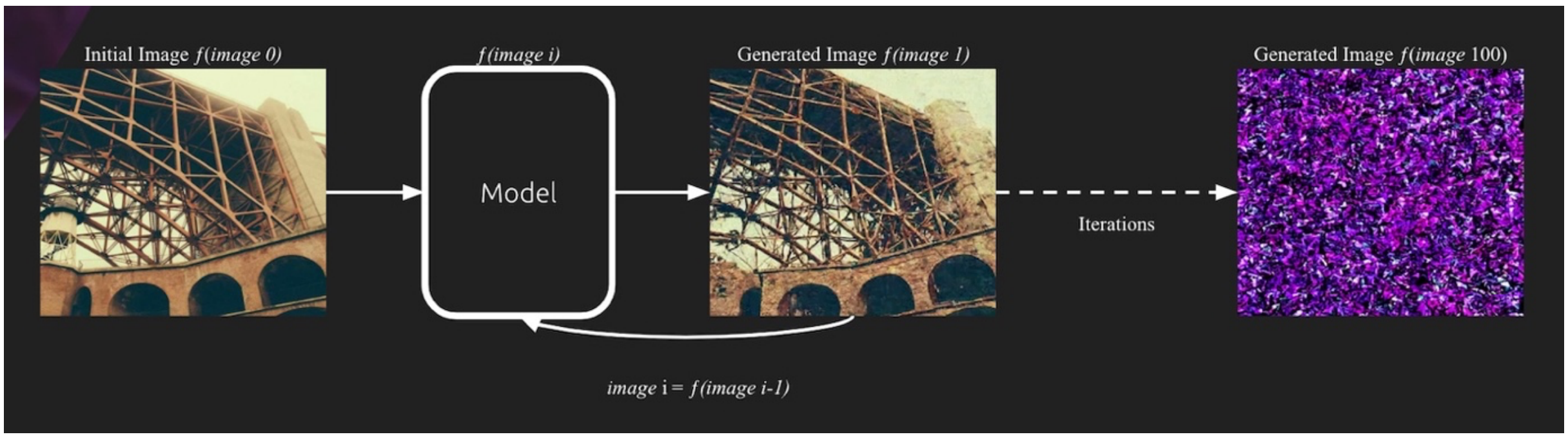

Figure 7: Ting-Chun Liu and Etienne-Leon Kühr, research data, 2024. The diagram shows Stable Diffusion outputs after the first and 100th iterations of recursive feeding, starting from a single image. Courtesy of the artists.

Like video feedback before it, this territory promises to be fertile yet calls for further artistic exploration. Meanwhile, one point remains clear: deep learning, for all its scale, still depends on continual access to vast quantities of high-quality human data, inherently noisier than synthetic ones. Without such external input, foundation models cannot maintain coherence on their own. In 1958, philosopher Gilbert Simondon defined the perfection of machines by their *margin of indeterminacy* (Simondon 1958, p. 11). Understood in media art as a condition that enables difference within repetition, a parallel can be drawn from an engineering perspective: human-generated content constitutes the margin of indeterminacy required for generative AI to sustain meaningful variation.

Nevertheless, beyond efficiency, model collapse invites us to reconsider generative AI not merely as a tool for processing perceived reality – or, more precisely again, our shared, digitized, or digital-native imagination – but as a powerful medium capable of unfolding entirely new worlds as well. It achieves this by turning our own inside out, revealing patterns previously unseen – some echoing earlier experiments with video feedback, others yet to be charted.

When China's DeepSeek model was described as a "Sputnik Moment for AI" (Acemoglu, 2025), it underscored how frontier models are increasingly bound up with state power, extra-terrestrial ambitions, and the rhetoric of space exploration. Interestingly, since the late 2000s theoretical physicists have used error-correcting codes – often involving recursive structures – to describe the universe, particularly in relation to supersymmetry (Doran, et al., 2008). Artists, too, have long turned to recursive techniques – from the golden ratio to Fibonacci spirals – to apprehend the infinite. Studying recursion not only as an agential entanglement but also as a formal and processual principle may thus help us understand how art reimagines perception and, perhaps, even touches the very fabric of the cosmos. This leaves us with one speculative question: might model collapse – and, more broadly, recursion and noise – become a site where the logics of artmaking and world-making converge?[9]

[9] The research underpinning the conclusion is currently supported by the Bibliotheca Hertziana – Max Planck Institute for Art History [BH-P-25-36].